\documentclass[%
 reprint,
superscriptaddress,
showpacs,preprintnumbers,
nofootinbib,
 amsmath,amssymb, 
 aps,
 prd,
 longbibliography,
]{revtex4-1}

\usepackage{cancel}
\usepackage{comment}
\usepackage{accents}
\usepackage{mciteplus,slashed}
\usepackage{amssymb,cancel,amsmath}
\usepackage{dcolumn}% Align table columns on decimal point
\usepackage{bm}% bold math
\usepackage[caption=false]{subfig}
\usepackage{appendix}
\usepackage{physics}
\usepackage{booktabs}
\usepackage{feynmp-auto}
\usepackage[T1]{fontenc}	
\usepackage{csvsimple}
\usepackage[bookmarksnumbered=true]{hyperref} 
\hypersetup{
    colorlinks = true,
    linkcolor = blue,
    anchorcolor = blue,
    citecolor = blue,
    filecolor = blue,
    urlcolor = blue
    }
\usepackage[section]{placeins}
\usepackage[capitalise]{cleveref}
\usepackage{booktabs}
\usepackage{graphicx}
\graphicspath{{Figures/}}

\AtBeginDocument{
\heavyrulewidth=.08em
\lightrulewidth=.05em
\cmidrulewidth=.03em
\belowrulesep=.65ex
\belowbottomsep=0pt
\aboverulesep=.4ex
\abovetopsep=0pt
\cmidrulesep=\doublerulesep
\cmidrulekern=.5em
\defaultaddspace=.5em
}
\usepackage[dvipsnames]{xcolor}
\usepackage[normalem]{ulem}
\usepackage{fontawesome} % for code link icons  

\definecolor{Red}{rgb}{1.,0.,0.}
\definecolor{Grn}{rgb}{0.,0.75,0.}
\definecolor{Blu}{rgb}{0.,0.,1.}
\definecolor{Purp}{rgb}{0.3,0.1,0.6}

\begin{document}

\title{Searching for new physics at $\mu\rightarrow e$ facilities with $\mu^+$ and $\pi^+$ decays at rest }
\author{Richard J.\ Hill}

\affiliation{Department of Physics and Astronomy, University of Kentucky,  Lexington, KY 40506, USA}
\affiliation{Theoretical Physics Department, Fermilab, Batavia, IL 60510,USA}

\author{Ryan Plestid}
\affiliation{Walter
 Burke Institute for Theoretical Physics, California Institute of Technology, Pasadena, CA 91125}
\affiliation{Department of Physics and Astronomy, University of Kentucky,  Lexington, KY 40506, USA}
\affiliation{Theoretical Physics Department, Fermilab, Batavia, IL 60510,USA}

\author{Jure Zupan}
\affiliation{Department of Physics, University of Cincinnati, Cincinnati, Ohio 45221, USA}

\preprint{FERMILAB-PUB-23-287-T}
\preprint{CALT-TH/2023-017}

\begin{abstract}
 We investigate the ability of $\mu\rightarrow e$ facilities, Mu2e and COMET, to probe, or discover, new physics with their detector validation datasets. The validation of the detector response may be performed  using a dedicated run with $\mu^+$, collecting data below the Michel edge, $E_e\lesssim 52$ MeV; an alternative strategy using $\pi^+\rightarrow e^+ \nu_e$ may also be considered.  We focus primarily on a search for a monoenergetic $e^+$ produced via two-body decays $\mu^+ \rightarrow e^+ X$ or $\pi^+\rightarrow e^+X$, with $X$ a light new physics particle. Mu2e can potentially explore new parameter space beyond present astrophysical and laboratory constraints for a set of well motivated models including:  axion like particles with flavor violating couplings ($\mu^+ \rightarrow e^+ a$), massive $Z'$ bosons ($\mu^+ \rightarrow Z' e^+$), and heavy neutral leptons ($\pi^+\rightarrow e^+N$).  The projected sensitivities presented herein  can be achieved in a matter of days. 
\end{abstract}

 \maketitle 
 %\tableofcontents
 \section{Introduction \label{Introduction}}
 
Charged lepton flavor violation (CLFV) is a long sought-after target of searches for physics Beyond the Standard Model (BSM) \cite{Feinberg:1958zzb,Minkowski:1977sc,Marciano:1977cj,Dzhilkibaev:1989zb,Czarnecki:1998iz,Kuno:1999jp,Sher:2002ew,Kitano:2002mt,BaBar:2005snc,Belle:2007diw,Raidal:2008jk,Marciano:2008zz,Proceedings:2012ulb,Bernstein:2013hba,deGouvea:2013zba,Gorringe:2015cma}. The most stringent limits come from searches for $\mu \rightarrow e \gamma$ \cite{MEGA:1999mhl,MEG:2016leq}, $\mu \rightarrow 3 e$ \cite{Bolton:1984qr,Bolton:1988af,SINDRUM:1985vbg,SINDRUM:1987nra}, and $\mu A \rightarrow e A$ transitions \cite{SINDRUMII:1996fti,SINDRUMII:2006dvw} (additional constraints arise from bounds on $\mu^- \rightarrow e^+$ conversion \cite{SINDRUMII:1998mwd}, muonium anti-muonium oscillations \cite{Willmann:1998gd,Conlin:2020veq}, and CLFV reactions involving $\tau$ leptons \cite{BaBar:2006jhm,Belle:2007cio,Arganda:2008jj,Gonderinger:2010yn,Celis:2013xja,Celis:2014asa,Takeuchi:2017btl,Cirigliano:2021img}). The $\mu A \rightarrow e A$ channel, often termed $\mu\rightarrow e$ or muon conversion, relies on the target nucleus to absorb recoil momentum, giving a kinematically allowed transition. Furthermore, if new physics mediating the $\mu \to e$ CLFV transition couples directly to quarks, then the presence of the nucleus itself catalyzes the reaction. 

Two upcoming facilities, COMET \cite{Kuno:2013mha,COMET:2018auw} and Mu2e \cite{Mu2e:2008sio,Mu2e:2014fns,Bernstein:2019fyh}, will search for $\mu \rightarrow e$ with unprecedented sensitivity -- the single-event sensitivities are expected to be as low as ${\rm BR}(\mu \rightarrow e)\sim 10^{-17}$. Both experiments leverage the extreme kinematics in $\mu \rightarrow e$, where almost all of the muon's rest mass is converted into the electron's kinetic energy.  The experiments therefore focus on the near endpoint region of maximal electron energy where  the Standard Model (SM) backgrounds are highly suppressed. Unfortunately, the same kinematic suppression applies to almost \emph{any} process other than $\mu \rightarrow e$, making searches for additional BSM decays using the high energy region datasets at Mu2e and COMET extremely challenging \cite{GarciaTormo:2011jit,Uesaka:2020okd}. 

In contrast, signal yields improve dramatically for many BSM scenarios 
in the regime of electron energy that is kinematically allowed for a free muon decay at rest. In this regime any particle lighter than the muon can be produced and  discovered with indirect search techniques. The simplest scenario to test is the two-body decay $\mu^+\rightarrow e^+X$, with $X$ the new light particle. The positively charged muon will decay at rest, resulting in a  monoenergetic positron signal. Both Mu2e and COMET are capable of collecting substantial $\mu^+$ (and $\pi^+$) datasets in this energy regime, which may be used for calibrating their detectors \cite{Private_comms_Pasha}. These datasets would have extremely high statistics relative to past $\pi^+$ and $\mu^+$ decay at rest searches \cite{shihua_thesis}, and are therefore well suited to search for light new physics. 

The Mu2e detector is designed to be charge symmetric such that both electrons and positrons can be reconstructed with high efficiency \cite{Bernstein:2019fyh}. Moreover, the design of the transport solenoid makes it possible to transport either $\mu^-$ or $\mu^+$ to the detector. COMET can also deliver $\mu^+$ on target \cite{Lee:2018wcx}. The use of $\mu^+$ decays instead of $\mu^-$ decays for  calibration has several advantages that also help enable a BSM search. 
Decays of $\mu^-$ are complicated by non-perturbative bound-state effects~\cite{Haenggi:1974hp,Watanabe:1987su,Czarnecki:2011mx,Czarnecki:2014cxa}
and backgrounds from radiative muon capture on the nucleus~\cite{Balashov:1967tgq,Fearing:1966zz,Eramzhian:1986aa}. 
In contrast, the Michel spectrum of $\mu^+\to e^+\nu\bar \nu$ decays is extremely well known, since it can be computed using standard diagrammatic techniques~\cite{Kinoshita:1958ru,Arbuzov:2002pp,Arbuzov:2002cn,Arbuzov:2002rp,Anastasiou:2005pn,Tomalak:2021lif,Banerjee:2022nbr}. Furthermore, the above-mentioned nuclear backgrounds are also mitigated due to the absence of muon capture for $\mu^+$. 

Note that such validation datasets can be used to search for
\emph{any} process that produces electrons or positrons close to the Michel edge. 
Important 
examples are the already mentioned two body $\mu^+ \rightarrow e^+ X$ decays, which result in monoenergetic positrons, but one could also search for non-standard multibody decays, where $X$ consist of several on-shell or off-shell new physics particles. 
A particularly interesting case is when $X$ is the QCD axion.  Our study shows that the Mu2e validation data can probe the region of parameter space in which the QCD axion is a cold dark matter candidate, assuming it has unsuppressed flavor violating couplings to muons and electrons~\cite{Calibbi:2020jvd}. 

At both COMET and Mu2e, the transport solenoid necessarily delivers $\pi^+$ along with $\mu^+$ to the target foils \cite{Lee:2018wcx,Bernstein:2019fyh}. The $\pi^+$ decay much faster than $\mu^+$, and can be separated with timing information and analyzed separately \cite{Bernstein:2019fyh}. 
In addition to non-standard muon decays, the large $\pi^+$ population in the validation dataset also enables a search for non-standard $\pi^+$ decays.  A phenomenologically important channel is the $\pi^+\rightarrow e^+ N$ decay, where $N$ is a heavy neutral lepton (HNL).

Motivated by its potential physics impact, we will use ``Mu2e-X'' as a shorthand for employing the Mu2e validation data for BSM searches, and similarly ``COMET-X'' for COMET. Mu2e is investigating the projected sensitivity of such a dataset internally \cite{Mu2e:2023aaa}. The rest of the paper is organized as follows: In \cref{UV} we outline new physics models, and regions of parameter space, that predict rates of $\mu^+\rightarrow e^+ X$ to which Mu2e will be sensitive.  We translate bounds on branching ratios to constraints on new physics model parameters, emphasizing the competitive reach of a $\mu^+\rightarrow e^+ X$ search relative to  astrophysical constraints. 
In \cref{sec:proj} we briefly describe
the inputs and procedures underlying 
our sensitivity estimates for $\mu^+\rightarrow e^+ X$ and $\pi^+ \rightarrow e^+ X$ searches. Finally, in \cref{concl} we summarize our findings and comment on possible future applications for the Mu2e validation data.

\section{Models of new physics \label{UV}}
We begin by discussing the theoretical motivation to search for two body decays $\mu^+ \rightarrow e^+ X$ and $\pi^+ \rightarrow e^+ X$. These are experimentally convenient because 
the predicted signature involves a monoenergetic $e^+$. Models with three body decays are also of interest but their experimental projections require further study; we briefly discuss this case in \cref{concl}.

In what follows we consider several benchmark new physics models for which a $\mu^+$ run at Mu2e could lead to a discovery or interesting limits.
Axion like particles (ALPs) can be discovered through two body $\mu^+\to e^+ X$ decays, while MeV scale DM can be searched for either through two body or three body $\mu^+\to (e^++{\rm invisible})$ decays. The rare $\pi^+\to e^+X$ decay mode can probe heavy neutral leptons (HNLs), but must overcome a sizable muon decay in flight background.

\subsection{Axion-like models \label{ALP}} 
Any spontaneously broken (approximate) global $U(1)$ symmetry results in a light (pseudo) Nambu-Goldstone boson in the low energy effective theory of the system. 
A particularly important example is the case of a spontaneously broken Peccei Quinn (PQ) symmetry giving rise to the QCD axion that can solve the strong CP problem and provide a cold dark matter candidate \cite{Peccei:1977ur,Weinberg:1977ma,Preskill:1982cy,Turner:1985si}. Such particles extending the SM are generically referred to as ALPs (axion like particles) \cite{DiLuzio:2020wdo}. 
If the spontaneously broken $U(1)$ is flavor non-universal, it can lead to sizable $\mu \to e a$ decays for ALPs $a$ with masses $m_a<105$ MeV \cite{Calibbi:2020jvd,Calibbi:2016hwq,Ema:2016ops,Choi:2017gpf,Arias-Aragon:2017eww,Linster:2018avp,Bjorkeroth:2018dzu,Bjorkeroth:2018ipq,Carone:2019lfc,Bauer:2019gfk,Bonnefoy:2019lsn,Cornella:2019uxs,Han:2020dwo,Escribano:2020wua,Bauer:2021mvw,DEramo:2021usm,Jho:2022snj,DiLuzio:2023ndz,Ziegler:2023aoe}.  

\begin{figure}
    \includegraphics[width=\linewidth]{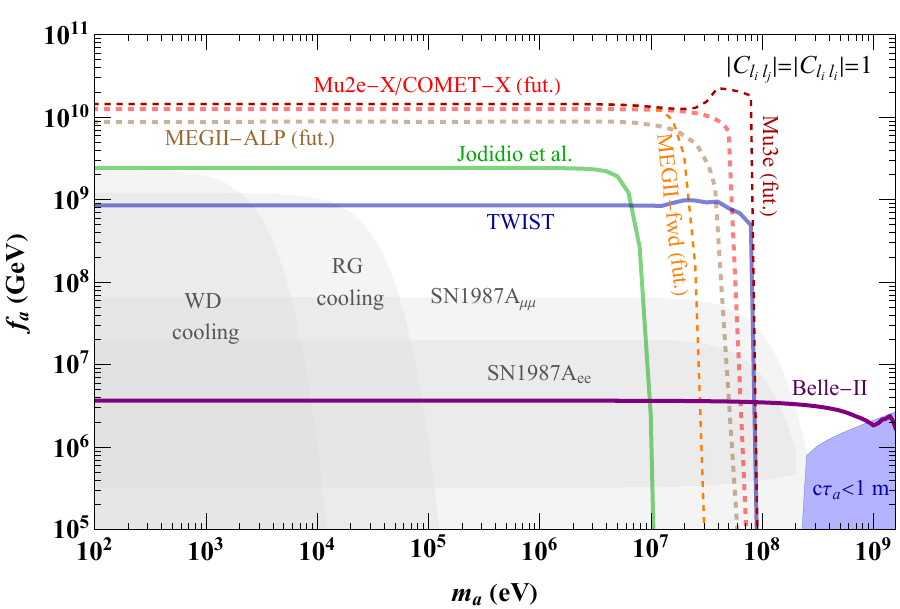}
    \caption{The 95\% C.L. limits on a general ALP with anarchic couplings to all three generations of leptons. The present laboratory exclusions are denoted with solid lines, and future projections with dashed lines, assuming isotropic ALP production with axial couplings, see text for further details. Astrophysical constraints are shown as gray region, while the parameter space that could lead to displaced decays inside detector volume, $c\tau_a<1\,$m is shown as blue region. Adapted and updated from Ref. \cite{Calibbi:2020jvd}. \label{fig:ALP:overview}}
\end{figure}

To understand whether or not a $\mu^+$ validation run could be sensitive to an interesting region of parameter space we explore three ALP benchmarks:  i) a general ALP with anarchic couplings to leptons (i.e., all couplings to leptons are of similar size), \cref{fig:ALP:overview},  ii) a leptophilic ALP that can be a DM candidate, \cref{fig:ALP:DM}, and iii) the QCD axion with lepton flavor violating couplings, \cref{fig:LFV:axion}.  The three benchmarks, along with other ALP models, were recently discussed in detail in Ref. \cite{Calibbi:2020jvd}.  Here we focus on the part of the phenomenology most relevant for $\mu^+\rightarrow e^+ X$. 

The effective Lagrangian describing the ALP couplings to the SM leptons ($\ell_i$)
%fermions,
%$f_i$, 
gluons ($G_{\mu\nu}$) and photons ($F_{\mu\nu}$) is given by\footnote{Note that $C_{ii}^V$ couplings do not contribute, as can be seen from equations of motion.}
\begin{equation}
\begin{split}
\label{eq:La:QCDaxion}
{\cal L}_a  =& \   N_{\rm UV} \frac{\alpha_s}{8 \pi} \frac{a}{f_a} G_{\mu \nu} \tilde{G}^{\mu \nu} + E_{\rm UV}  \frac{\alpha_{\rm em}}{4 \pi} \frac{a}{f_a} F_{\mu \nu} \tilde{F}^{\mu \nu}  
\\
&+\sum_{i,j} \frac{\partial_\mu a}{2 f_a} \overline{\ell}_i \gamma^\mu \left[ C^V_{ij} + C^A_{ij} \gamma_5 \right] \ell_j~,
\end{split}
\end{equation}
where $i,j=1,2,3$ are generational indices, color indices are suppressed, and the subscript UV denotes ``ultraviolet''. Since we are mostly interested in processes involving leptons, the equivalent couplings to quarks are set to zero. The derivative couplings are a hallmark of the pseudo Nambu-Goldstone boson (pNGB) nature of the ALP, i.e.,\ we assume that the shift symmetry is softly broken only by the ALP mass, $m_a$. All the couplings in \eqref{eq:La:QCDaxion} are of dimension 5 and are suppressed by the ALP decay constant, $f_a$, which can be identified with the scale of spontaneous symmetry breaking. 

For $i\ne j$, the ALP couplings are flavor violating. In new physics models with no particular flavor structure the generic expectation would be that $C_{ij}^{V/A}$ are all nonzero. If this is the case, the flavor changing neutral current (FCNC) constraints, either from $\mu\to e a$, or from $K\to \pi a$ decays in the case of couplings to quarks,  impose very stringent constraints,  $f_a\gtrsim 10^{9}$ GeV and $f_a\gtrsim 10^{12}$ GeV when assuming ${\mathcal O}(1)$ flavor violating couplings to either leptons or  quarks, respectively \cite{Calibbi:2020jvd,MartinCamalich:2020dfe}. The sensitivity of $\mu\to e a$ to such high scales can be traced to the fact that on-shell production of an ALP is induced by dimension 5 operators, and thus ${\rm BR}(\mu \to e a)\propto (m_\mu/f_a)^2$. This can be contrasted with the much weaker constraints on such models from $\mu\to e$ conversion~\cite{Fuyuto:2023ogq}, which require two insertions of dimension 5 operators (the flavor violating coupling to leptons and the flavor conserving coupling to quarks), giving ${\rm BR}(\mu\to e)\propto (m_\mu/f_a)^4$, i.e., a rate that is additionally suppressed by a factor $(m_\mu/f_a)^2\sim 10^{-20}$ compared to ${\rm BR}(\mu \to e a)$.

 For quantitative analysis we first consider three benchmarks from Ref. \cite{Calibbi:2020jvd}, and then discuss implications for other ALP models: 
 %first, an ALP with anarchic couplings to leptons; second, a leptophilic ALP as a DM candidate; and finally, a lepton flavor violating QCD axion, and then discuss other possible ALP scenarios.

%%%%%%%%%%%%%%%%%%%%%%%%%%%%%%%%%%%%%%%%%%%%%%%
\subsubsection{ALP with anarchic couplings to leptons}
%%%%%%%%%%%%%%%%%%%%%%%%%%%%%%%%%%%%%%%%%%%%%%%% 
In the first benchmark case, the ALP is assumed to couple only to  leptons with both flavor violating and flavor conserving couplings of similar size. For concreteness we assume that all axial couplings to lepton are the same and equal to $C_{ij}^{A}=1$, the vector couplings are assumed to vanish, $C_{ij}^{V}=0$, as do the direct couplings to photons and gluons, i.e., we set $E_{\rm UV}=N_{\rm UV}=0$. The couplings to photons (gluons) are still generated radiatively at one loop (two loops) from couplings to leptons, but are not relevant for phenomenological studies. The ALP mass, $m_a$, is treated as a free parameter. The projected $95\%$ C.L.\ constraints on this benchmark are shown in \cref{fig:ALP:overview} with red dashed line.\!\footnote{Here we appropriately rescale the results for $90\%$ CL bounds from \cref{sec:proj} to 95\% CL interval, which all the shown bounds use.}

The present laboratory constrains are shown with solid green \cite{Jodidio:1986mz}  and blue \cite{TWIST:2014ymv}. These constraints depend on the chiral structure of the ALP couplings, and are, for instance, significantly relaxed for $V-A$ couplings in the case of constraints from Ref.~\cite{Jodidio:1986mz}.  The present constrains from $\tau\to \ell a$ decays are shown with a solid  purple line \cite{Belle-II:2022heu}, while the astrophysics constraints are shown as gray excluded regions; see Ref.~\cite{Calibbi:2020jvd} for further details. In \cref{fig:ALP:overview} we show with dashed orange and dark red curves the future sensitivities at MEGII-fwd, assuming realistic focusing \cite{Calibbi:2020jvd}, and the projected sensitivity at Mu3e \cite{Perrevoort:2018okj}, respectively.
A similar reach in $f_a$ could be also achieved by searching for $\mu\to e a \gamma$ decays at MEG-II after one year of running in an alternative data taking strategy with reduced beam intensity and adjusted triggers \cite{Jho:2022snj}, shown with brown dashed line (we show the upper range of the projected sensitivity band in \cite{Jho:2022snj}).
We see that Mu2e-X and COMET-X have comparable reach to these other proposals to search for $\mu\to e a$ transitions. 

%%%%%%%%%%%%%%%%%%%%%%%%%%%%%%%%%%%%%%%%%%%%%%%
\subsubsection{Leptophilic ALP as a DM candidate}
\begin{figure}
    \includegraphics[width=\linewidth]{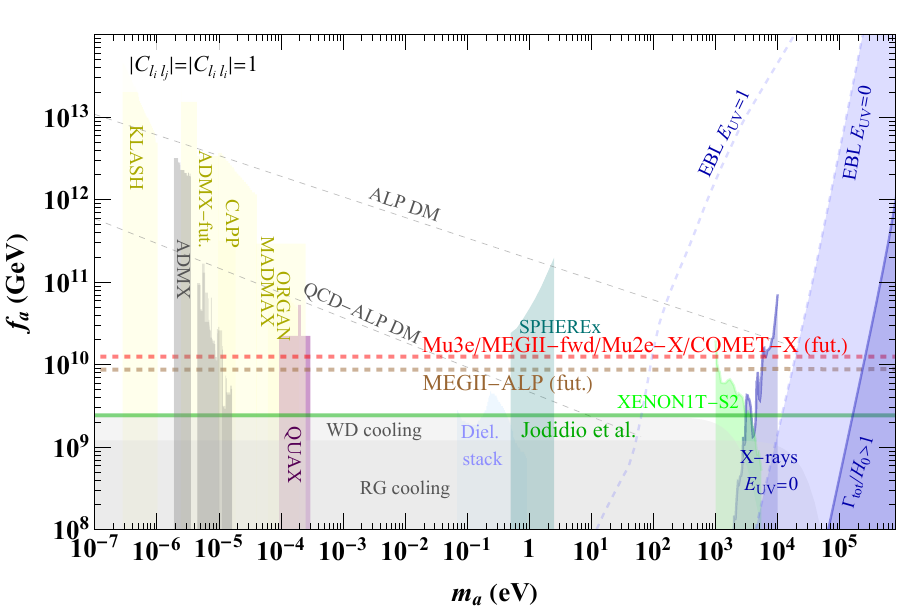}
    \caption{The 95\% C.L. limits on a leptophilic ALP that can be a DM candidate, as well as the reach of a $\mu^+$ run (red dashed line, labeled Mu2e-X), see main text for details. Mu2e-X, COMET-X, MEGII-fwd, and Mu3e have similar projected sensitivities, and we represent all of them with a single line.
    Adapted from Ref.~\cite{Calibbi:2020jvd}.
    % \rp{Update lines}. 
    \label{fig:ALP:DM}}
\end{figure}

%%%%%%%%%%%%%%%%%%%%%%%%%%%%%%%%%%%%%%%%%%%%%%%
If an ALP is light enough it becomes cosmologically stable and can be a DM candidate. \cref{fig:ALP:DM} shows the constraints for such a possibility with anarchic couplings to leptons, $C_{ij}^{V/A}=1$, and no direct couplings to gluons $N_{\rm UV}=0$. The constraints from extragalactic background light bounds are shown for two cases $E_{\rm UV}=1$ (dashed blue line) and $E_{\rm UV}=0$ (light blue region), where regions to the right are excluded. The ALP DM (QCD-ALP DM) dashed line shows the parameter space for which the initial misalignment of the ALP field in the early universe, $\theta_0=1$, leads to the correct relic DM abundance, assuming no temperature corrections to the ALP mass (i.e., thermal mass dependence equivalent to that of the QCD axion).

The green solid line in \cref{fig:ALP:DM} shows the current best bound on the isotropic LFV ALP~\cite{Jodidio:1986mz}, the brown dashed line denotes the most optimistic projected reach from $\mu\to e a \gamma$ decays at MEG-II after one year of running, while the red dashed line shows the expected reach of Mu2e, which is comparable to the MEGII-fwd projection including focusing enhancement and Mu3e. The expected reach is well above the existing and future bounds that rely on couplings between ALPs and electrons, shown as color shaded regions, and can probe parameter space where the flavor violating ALP is a viable DM candidate.

\begin{figure}
    \includegraphics[width=\linewidth]{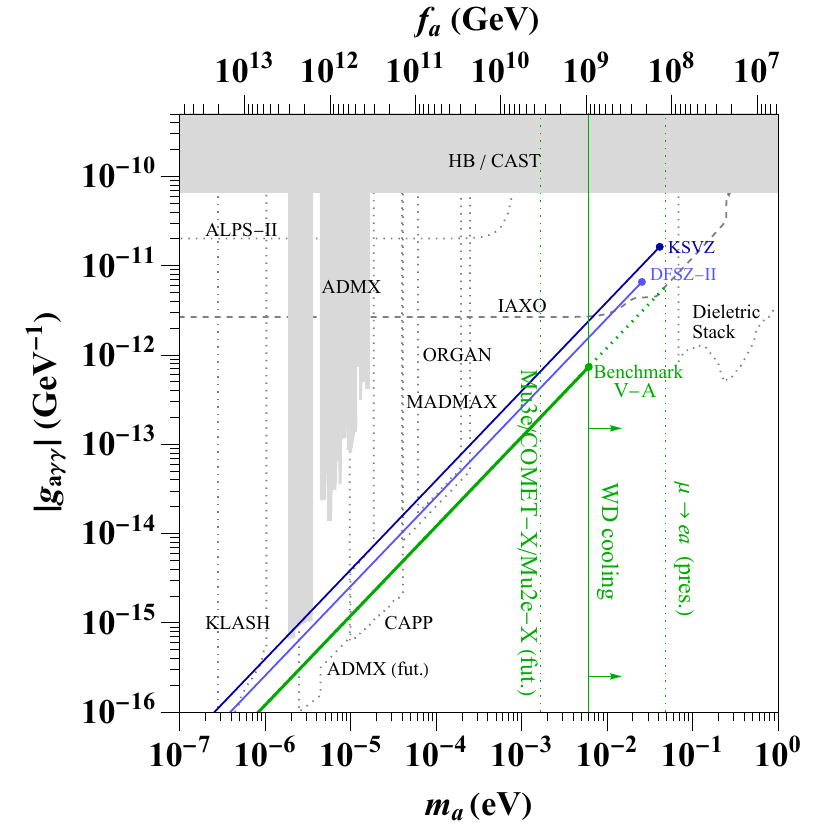}
    \caption{The 95 \% C.L. limits on lepton flavor violating QCD axion for the assumed $V-A$ forms of couplings. The mass of the QCD axion, $m_a$, is inversely proportional to the coupling constant $f_a$.
    The vertical axis refers to the axion coupling to photons, $g_{a\gamma\gamma}\propto \alpha_{\rm em} /(2 \pi f_a)$, where an additional constant coefficient depends on the particular model. The benchmark $V-A$ LFV QCD axion model is indicated by the tilted solid green line. Also shown are two other QCD axion models not involving LFV, the KSVZ~\cite{kim1979weak,shifman1980can} (dark blue) and the DFSZ-II (blue) model, having slightly different couplings to photons. 
   The current excluded ranges of $g_{a\gamma\gamma}$ as function of $m_a$ are shown as shaded gray regions, and future projected limits as dashed gray lines. The  $\mu\to e a$ limits, which are independent of $g_{a\gamma\gamma}$, but assume sizable LFV couplings, exclude values of $m_a$ to the right of the dashed vertical green lines under these assumptions (but thus do not apply to KSVZ and DFSZ-II models). The solid green vertical line refers to the limit from white dwarf (WD) cooling constraint which assumes sizable axion coupling to electrons. The sensitivity derived from Mu2e calibration data (dotted vertical green line) will probe parameter space beyond this limit. Adapted from Ref.~\cite{Calibbi:2020jvd}. 
   %\rp{Update lines} 
   \label{fig:LFV:axion}}
\end{figure}
The relevant space in \cref{fig:ALP:DM} is to the left of the blue region enclosed by the solid blue line, which delineates the parameter space leading to ALP decaying within the present Hubble time. The region to the right of the dashed blue lines is excluded by the extragalactic diffuse background light measurements for $E_{\rm UV}=0,1$, as denoted. The dark blue region shows the X-rays constraints for $E_{\text{UV}}=0$~\cite{Boyarsky:2006hr,Figueroa-Feliciano:2015gwa}. The gray shaded regions are excluded by the star cooling bounds, and the ADMX results~\cite{Braine:2019fqb,ADMX:2018ogs,ADMX:2018gho}. The light green region is excluded by the S2 only analysis of XENON1T~\cite{XENON:2019gfn} and Panda-X~\cite{PandaX:2017ock}. The purple shaded region shows the future reach of axion-magnon conversion experiment QUAX~\cite{Barbieri:1985cp,Barbieri:2016vwg,Chigusa:2020gfs}.  The cyan colored region shows the future sensitivity of SPHEREx experiment that relies on ALP couplings to photons, assuming ALP decays exclusively to two photons~\cite{Creque-Sarbinowski:2018ebl}, while the yellow regions show the future sensitivities of resonant microwave cavity searches: ADMX~\cite{Shokair:2014rna}, CAPP~\cite{Petrakou:2017epq}, KLASH~\cite{Gatti:2018ojx}, and ORGAN~\cite{McAllister:2017lkb}, as well as the searches using dielectric haloscope  MADMAX~\cite{Caldwell:2016dcw} or  (light blue region) using dielectric stacks~\cite{Baryakhtar:2018doz}. The $\mu^+\rightarrow e^+a$ limit using Mu2e-X is complementary to all these searches.

%%%%%%%%%%%%%%%%%%%%%%%%%%%%%%%%%%%%%%%%%%%%%%%
\subsubsection{Lepton flavor violating QCD axion}
%%%%%%%%%%%%%%%%%%%%%%%%%%%%%%%%%%%%%%%%%%%%%%%
Mu2e calibration data can also be sensitive to a QCD axion that solves the strong CP problem. The QCD axion will have flavor violating couplings, if the PQ symmetry is not flavor universal  \cite{Calibbi:2016hwq,Wilczek:1982rv}. The mass of such a flavor violating QCD axion still arises entirely from the QCD anomaly, $m_a = 5.691(51) \mu\text{eV} \big({10^{12} \,\text{GeV}}/{f_a} \big)$~\cite{Gorghetto:2018ocs}, and is thus effectively massless in $\mu \to e a$ decays. The flavor violating QCD axion is also a viable cold dark matter candidate.  If the axion relic abundance is due to the misalignment mechanism, the $\theta_0\sim {\mathcal O}(1)$ misalignment angle leads to the observed DM relic density for axion decay constants in the range $f_ a \sim 10^{(11 - 13)}$ GeV. For smaller decay constants, within the reach of LFV experiments, the axion relic from the standard misalignment contribution is under-abundant unless the relic abundance is due to some non-trivial dynamics.

In \cref{fig:LFV:axion} we show constraints on a particular DFSZ-like model~\cite{zhitnitskij1980possible,dine1981simple} of the QCD axion with LFV couplings~\cite{Calibbi:2020jvd} (tilted solid green line).  The field content of the theory consists of the SM fermions, two Higgs doublets, $H_{1,2}$, and a complex scalar $S$ that is a gauge singlet. 
The model contains an anomalous global $U(1)$ PQ symmetry under which all the scalars are charged. It is broken once $S$ obtains a vacuum expectation value (vev), giving rise to the light pNGB -- the QCD axion. The PQ charges of the SM leptons are generation dependent such that $H_2$ couples only to second and third generation leptons, while the $H_1$ lepton Yukawa interactions couple first generation to second and third generation leptons. The generation dependent PQ charges then translate to flavor violating axion couplings to leptons. The PQ charges of quarks are universal and thus the axion has flavor diagonal couplings to quarks.

 The constraints in \cref{fig:LFV:axion} are shown for a particular benchmark where the QCD axion couplings to the leptons have $V-A$ chiral structure, and where the flavor violating couplings involving $\tau$-leptons are assumed to be suppressed (see Ref.~\cite{Calibbi:2020jvd} for details). We see that the sensitivity obtainable at Mu2e-X will probe parameter space well beyond the present astrophysics bound from white dwarf cooling constraints (solid green line), improving on the present  laboratory bounds from searches for $\mu\to ea$ decays. The $\mu\to e a$ lines are vertical, since they are insensitive to the axion coupling to photons, $g_{a\gamma\gamma}= -0.59 \times \alpha_{\rm em} /(2 \pi f_a)$.

%\vspace{-12pt}
%%%%%%%%%%%%%%%%%%%%%%%%%%%%%%%%%%%%%%%%%%%%%%%
\subsubsection{Other possible ALP models}
%%%%%%%%%%%%%%%%%%%%%%%%%%%%%%%%%%%%%%%%%%%%%%%
The above examples by no means exhaust the set of possible models that could be searched for via $\mu \to e X$ decays. Importantly, the flavor structures of flavor violating couplings in the above three examples were fixed externally. In some models the pattern of flavor violating couplings is instead determined by the dynamics of the new physics model itself. An example is the ``axiflavon'' model,  in which the QCD axion $a$ is responsible both for generating the observed flavor structure of the SM as well for solving the strong CP problem \cite{Calibbi:2016hwq,Ema:2016ops}. The axiflavon is a representative of  an entire class of ``familon'' theories \cite{Wilczek:1982rv,Reiss:1982sq,Gelmini:1982zz,Chang:1987hz}  in which the ALP is associated with a spontaneously broken horizontal family symmetry, e.g.,\ of a Froggatt-Nielsen type \cite{Froggatt:1978nt} or from a nonabelian global horizontal group such as SU(2) \cite{Linster:2018avp}. In these scenarios a large $\mu-e$ CLFV coupling is \emph{predicted} such that a search for $\mu^+\rightarrow e^+X$ can test these models and offer an avenue to discovery (see recent work on testing at Mu2e such models with heavy familons \cite{Koltick:2021slp}). Here, we argue that Mu2e-X is in fact capable of probing important parameter space across a wide range of familon masses. 

The $\mu \to e X$ transition can also probe dynamical models of neutrino mass generation, where $X$ is the Majoron, a pNGB of  a spontaneously broken lepton number \cite{Chikashige:1980ui,Schechter:1981cv,Pilaftsis:1993af,Heeck:2019guh}. In TeV-scale see-saw mechanism the neutrino masses are parametrically suppressed while CLFV couplings are not \cite{Broncano:2002rw,Raidal:2004vt,Kersten:2007vk,Abada:2007ux,Gavela:2009cd,Ibarra:2010xw,Dinh:2012bp,Cely:2012bz,Alonso:2012ji}. The parametric suppression of neutrino masses is technically natural and can emerge from an approximate symmetry of a generalized lepton number $U_{L'}(1)$ under which the CLFV couplings are invariant, while the neutrino masses are not, and must be proportional to a small symmetry breaking parameter. This can then result in a potentially observable $\mu\to e X$ decays.

As outlined in \cite{Calibbi:2020jvd} and also shown in \cref{fig:ALP:overview,fig:LFV:axion}, the ability of laboratory experiments to probe branching ratios of ${\rm BR}(\mu \rightarrow e X) \lesssim 10^{-5}$ results in constraints on ALP couplings that for generic flavor structures supersede the already stringent bounds from astrophysical sources. This allows experiments such as Mu3e \cite{Perrevoort:2018okj}, MEGII-fwd \cite{Calibbi:2020jvd}, and, as we argue here, Mu2e, to provide leading constraints on ALP models.
% 

%%%%%%%%%%%%%%%%%%%%%%%%%%%%%%%%%%%%%%%%%%%%%%%
\subsection{Heavy Neutral Leptons \label{HNLs}}
%%%%%%%%%%%%%%%%%%%%%%%%%%%%%%%%%%%%%%%%%%%%%%%%

Models with heavy neutral leptons (HNL) \cite{Shrock:1980vy,Shrock:1980ct,Shrock:1981wq,Abela:1981nf,Asano:1981he,Hayano:1982wu,Ellis:1982ve,Minehart:1981fv,Azuelos:1986eg,Gorbunov:2007ak,Drewes:2013gca,Essig:2013lka,Ballett:2019bgd,Beacham:2019nyx,Lanfranchi:2020crw,Agrawal:2021dbo}, i.e.,\ sterile neutrinos with masses in the MeV to few GeV range, have received substantial attention over the past fifteen years in the context of light dark sectors \cite{deGouvea:2015euy,Cherry:2014xra,Bertuzzo:2018itn,Magill:2018jla,Bondarenko:2018ptm,Bolton:2019pcu,Bryman:2019bjg,Beacham:2019nyx,Plestid:2020ssy,Plestid:2020vqf,Coloma:2019htx,Hostert:2020xku,Agrawal:2021dbo,Abdullahi:2022jlv,Gustafson:2022rsz}. Couplings between HNLs, $N$ (with mass $m_N$), and SM neutrinos, $\nu$, offer one of three renomalizable ``portals'' between a dark sector and the SM \cite{Batell:2009di,Alexander:2016aln}, 
\begin{equation}
\label{eq:HNL}
{\cal L}\supset y_N(LH) N+{\rm h.c.},
\end{equation}
where $L$ is the SM lepton doublet, $H$ is the Higgs doublet, and we suppress flavor indices. After electroweak symmetry breaking the Yukawa interaction \eqref{eq:HNL} 
induces mixing between HNLs and SM neutrinos, through which dark sector degrees of freedom may imprint themselves on experimental data. For $\pi^+\rightarrow e^+N$ the relevant mixing parameter in the extended PMNS matrix~\cite{pontecorvo1957no,maki1962remarks} is $U_{eN}= \braket{N}{\nu_e}$. Searching for HNLs is of 
interest both for minimal and extended dark sectors  \cite{Beacham:2019nyx,Agrawal:2021dbo}. 
\begin{figure}
    \includegraphics[width=\linewidth]{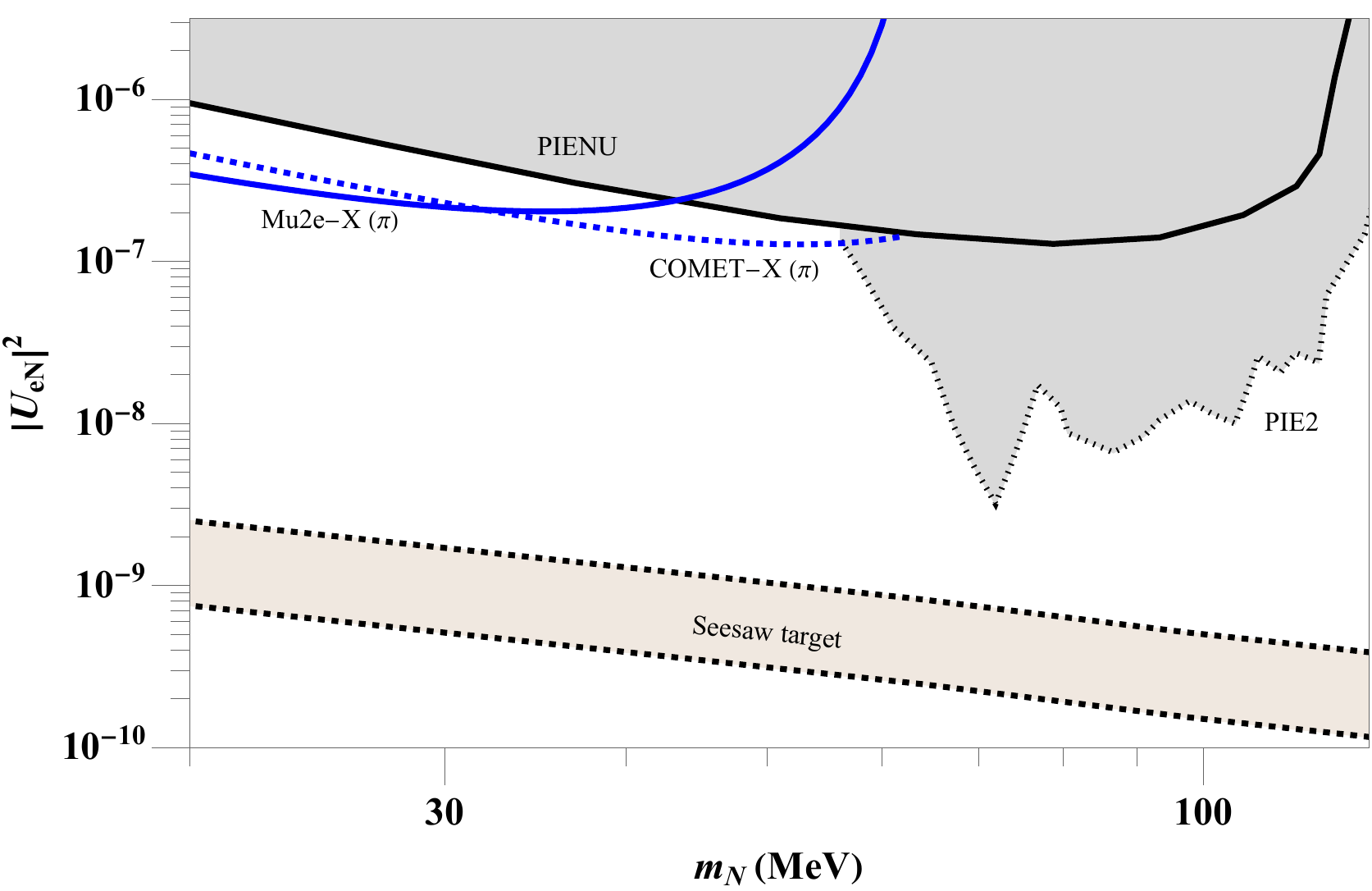}
    \caption{Projections for mass dependent 90\%-CL sensitivity to HNL mixing with electron flavor from Mu2e-X in a pion configuration, see \cref{sec:proj} for details. Existing limits come from PIENU \cite{PIENU:2017wbj,aguilar2015improved,Bryman:2019ssi} and their related bump hunt PIE2 \cite{aguilar2020improved} (see also \cite{Bryman:2019ssi} for a compilation). The different shapes at Mu2e and COMET arise due to the different model acceptances used in this analysis. 
    \label{fig:HNL}}
\end{figure}

In principle, either $\pi^+\rightarrow \ell^+ N$ or $\mu^+ \rightarrow e^+ N \nu$ decays can be used to search for HNLs, provided $N$ is light enough to be produced in these decays. Mu2e is both a muon and a pion factory, and large populations of both particles are delivered to the stopping target. The challenge in searching for $\mu^+ \rightarrow e^+ N \nu$  decays is that the background due to SM muon decays is also three body.  Fitting for the spectral distortion from the HNL in the observed Michel spectrum is in principle possible, but made more challenging by the complicated energy dependent acceptances in Mu2e due to the helical tracker. Furthermore, such a search would require a detailed understanding of background spectra and radiative corrections to the Michel spectrum.

Constraints on HNL models are conventionally studied in a single-flavor mixing paradigm with constraints appearing in the $m_N-|U_{\alpha N}|$ plane with $\alpha\in \{ e,\mu,\tau\}$ labeling the lepton flavor that the HNL couples to. In the case of pion decay to $e^+N$ the relevant parameter is $U_{e N}$, and the range of HNL masses that can be probed (in principle) is $m_N \in [0, m_{\pi}-m_e]$.  The branching ratio, for $m_e\ll m_N \lesssim m_\pi$, is given by
\begin{equation}
     {\rm BR}(\pi^+ \rightarrow e^+ N)= |U_{e N} |^2 \frac{m_N^2(m_\pi^2 -m_N^2)^2}{m_\mu^2(m_\pi^2 -m_\mu^2)^2} \,.
     \label{BR-HNL}
\end{equation}
The dominant background is due to $\mu^+ \rightarrow e^+ \nu \nu$ decays, which can be significantly reduced  with timing and geometric cuts \cite{shihua_thesis}. In \cref{fig:HNL} we show a compilation of limits from existing experiments and overlay projections for Mu2e-X in a configuration that could be used during a $\pi^+\rightarrow e^+ \nu$ calibration of the detector response. Since the sensitivity to an HNL is highly mass dependent, {\it cf.}  \cref{BR-HNL}, we focus on the region $m_N/m_\pi \lesssim {\mathcal O}(1)$, and leave the very light HNL mass range,  $m_N<20$ MeV for future dedicated studies.

%%%%%%%%%%%%%%%%%%%%%%%%%%%%%%%%%%%%%%%%%%%%%%%
\subsection{Dark $Z'$ models \label{dark-z'}}
%%%%%%%%%%%%%%%%%%%%%%%%%%%%%%%%%%%%%%%%%%%%%%%
Another class of models that may be searched for at Mu2e-X are the BSM models that contain a light flavor violating $Z'$ \cite{Smolkovic:2019jow,Ardu:2022zom,Ibarra:2021xyk,Foot:1994vd,Heeck:2016xkh,Alonso-Alvarez:2021ktn,Araki:2021vhy,Dev:2020drf,Heeck:2017xmg,Altmannshofer:2016brv}, decaying predominantly  to invisible final states, $Z'\to \chi \bar\chi$. Here,  $\chi$ can be dark matter or a mediator to the dark sector \cite{Baek:2008nz,Bauer:2018egk,Arcadi:2018tly}. The effective Lagrangian, assuming renormalizable interactions, is given by\footnote{The $\mu \to e Z'$ decays could in general also occur through dimension 5 dipole operators, see, e.g.,\ Ref.~\cite{Ibarra:2021xyk}~.} \cite{Smolkovic:2019jow}
\begin{equation}
\begin{split}
\label{eq:cfi:Z'}
{\mathcal L}
\supset & Z_\mu' g' \bar \chi \gamma^\mu \chi+ Z_\mu' \sum_{f,i,j} \Big[ g'c_{f_L}^{ij} \big(\bar f_L^{(i)}\gamma^\mu  f_L^{(j)}\big)
\\
&\hspace{0.4\linewidth}+g' c_{f_R}^{ij} \big(\bar f_R^{(i)}\gamma^\mu  f_R^{(j)}\big)\Big],
\end{split}
\end{equation}
where we assumed that $\chi$ is a Dirac fermion, while the sum runs over all the SM fermions, $f=u,d,\ell, \nu$, and the generation indices $i,j=1,\ldots,3,$. Assuming $\chi$ is light enough that $Z'\to \chi \bar \chi$ decays are kinematically allowed, these will dominate over the $Z'$ decays to SM fermions, as long as the corresponding effective coefficients are small, $|c_{f_{L,R}}^{ij}|\ll 1$. A concrete realization of such a scenario is a $Z'$ that is the gauge boson of a dark $U(1)_X$ under which the dark sector is charged, while the interactions of the SM fermions are induced through mixings with dark vector-like fermions. In general, this induces both flavor conserving and flavor violating couplings  $c_{f_L}^{ij}$, $c_{f_R}^{ij}$. The $\mu \to e Z'$ decay width is given by \cite{Smolkovic:2019jow}
\begin{equation}
\begin{split}
\Gamma(\mu \to eZ')= \Big[&(c_{\ell_L}^{12})^2+(c_{\ell_R}^{12})^2\Big] \\
&\times \frac{g'{}^2}{32\pi } \frac{m_\mu^3}{m_{Z'}^2}\big( 1-  r_{Z'}\big)^2 \big( 1 + 2r_{Z'} \big) \,,
\end{split}
\end{equation}
where we neglected the terms suppressed by $m_e/m_\mu$, and shortened $r_{Z'}={m_{Z'}^2}/{m_\mu^2}$. The mass of the $Z'$ gauge boson is given by $m_{Z'}=g' v'$, where $v'$ is the vev that breaks the $U(1)$ gauge symmetry. We see that $\Gamma(\mu \to eZ')\propto \Big[(c_{\ell_L}^{12})^2+(c_{\ell_R}^{12})^2\Big]/{v'^2}$, and is vanishingly small if either $c_{\ell_{L,R}}^{12}\to 0$, or if $v'$ is large.

\begin{figure}[t]
\includegraphics[width=0.999\columnwidth]{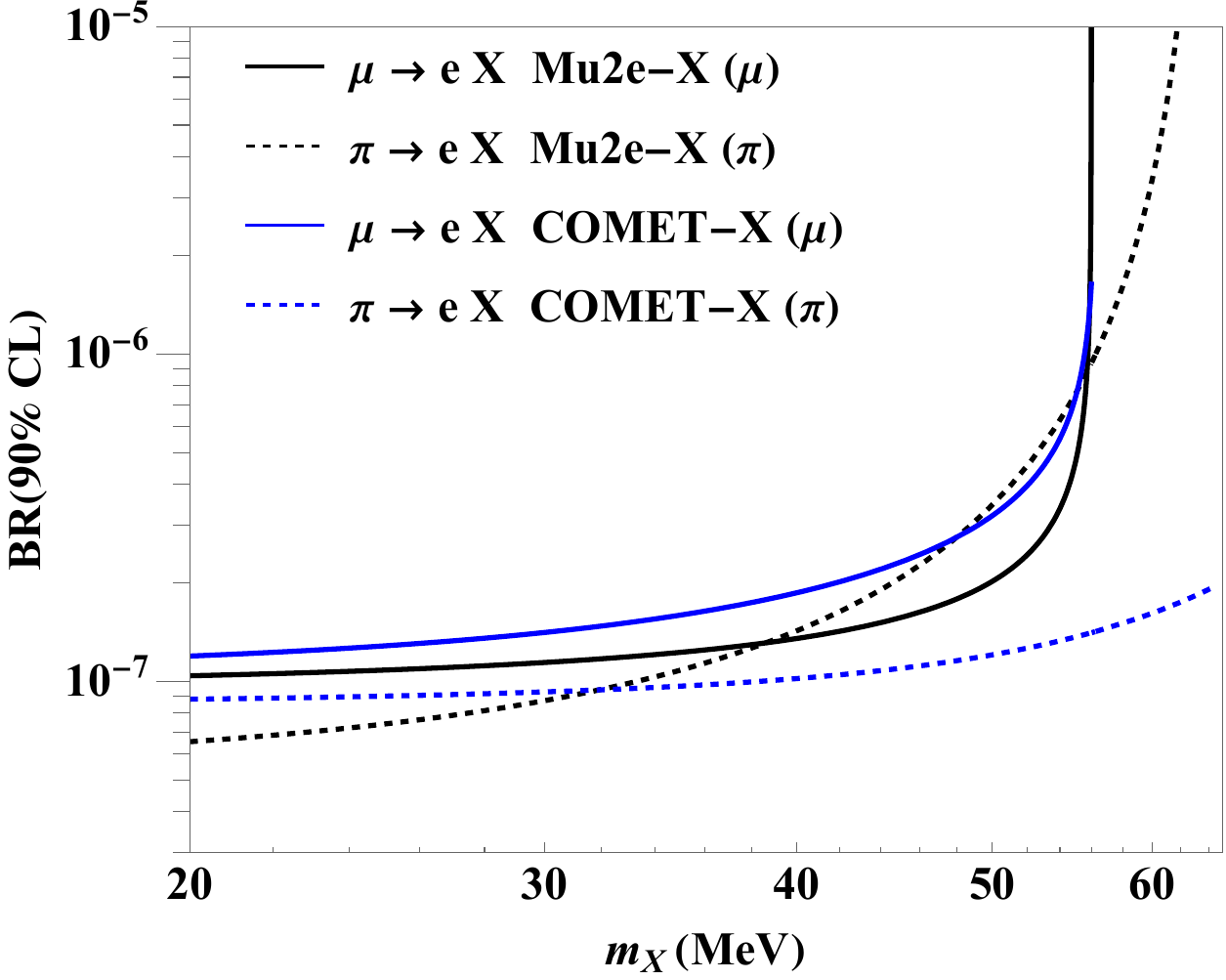}
\caption{\label{fig:sensitivity} Estimated branching ratio limits (90\% C.L) for $\mu\rightarrow eX$ and $\pi \rightarrow e X$ as a function of $m_{X}$ for Mu2e-X and COMET-X. The shape of the exclusion depends both on the acceptance as a function of energy, and on the background as a function of energy. See \cref{inputs} for details of the inputs used in estimating projected sensitivity. The different shapes of the COMET-X ($\pi$) and Mu2e-X ($\pi$) curves arise due to different acceptances which depend on the positron momentum.  }
\end{figure}

Another example of flavor violating light $Z'$ is the possibility that the $U(1)_X$ is the horizontal symmetry responsible for the hierarchy of SM fermion masses, such as in the Froggatt-Nielsen model of Ref.~\cite{Smolkovic:2019jow}. In that case the FCNC bounds from other states in the theory require $v'\gtrsim 10^{7}$ GeV, while $Z'$ can be light if $g'\ll 1$. Both invisible decays, $Z'\to \nu\bar\nu$, and visible decays to SM fermions $Z'\to f\bar f$ need to be considered in the final state since both can have large branching ratios. The values of $c_{f_{L,R}}^{ij}$ coefficients depend on the details of the numerical inputs in the model benchmark, but are in general ${\mathcal O}(1)$ for diagonal and $10^{-3}$--$10^{-1}$ for off-diagonal entries \cite{Smolkovic:2019jow}. 

Let us close this section by mentioning the possibility of neutrino-induced CLFV couplings due to heavy neutral leptons. Models of this type have been studied in the context of neutrino portal dark matter~\cite{Blennow:2019fhy,Batell:2017cmf}, and produce off-diagonal CLFV couplings via triangle diagrams, with flavor mixing from an (extended) PMNS matrix. The result is an off-diagonal flavor coupling given by ({\it cf.} Eq.~(6.4) of~\cite{Blennow:2019fhy})
\begin{equation}
    c^{ij}_{L,R} \approx U_{iN} U_{j N}^* \frac{g^2}{4\pi} \frac{m_N^2}{m_W^2}~,
\end{equation}
where $U_{iN}$ are the PMNS matrix elements between flavor $i$ and the HNL, $N$, and we have assumed that $N$ is very nearly aligned with the mass basis. 

The search strategy we propose is model independent, relying only on the two-body final state kinematics. Unlike axion models, which are generically expected to be very light due to the approximate global $U(1)$ symmetry, in many $Z'$ scenarios the dark vector is massive ($m_{Z'}\gtrsim 10$ MeV) to avoid BBN~\cite{Workman:2022ynf} bounds. A massive $Z'$ furnishes a theoretically well motivated candidate with $m_X>20$ MeV that can be searched for in the Mu2e validation data. 

\begin{table}[t]
\caption{%
\label{inputs}
Parameters used to estimate sensitivities in this work: the number of stopped parents, $P\in \{ \mu ,\pi\}$, the assumed operating $B$-field relative to nominal, $B/B_0$, the number of background events expected in the signal region, $\mu_{\rm bkg}$, and the efficiency/acceptance $\epsilon_P$ as a function of $P_{e^+}$.  The expected number of background events for the $(\mu)$-configuration are estimated using the tree-level Michel spectrum, whereas in a $(\pi)$-configuration they come from muon decay in flight~\cite{shihua_thesis}.  }
\begin{ruledtabular}
\begin{tabular}{lccccc}%{lcdr}
\textrm{Configuration}& \textrm{$N_P$}&  $B/B_0$ &$\mu_{\rm bkg}$ & $\epsilon_P(P_{e^+})$\\
\colrule
Mu2e-X ($\mu$) &  $3\times 10^{13}$ &  $0.5$ & \cref{mu-bkg} & \cref{eff-mu} \\
Mu2e-X ($\pi$)&  $2\times 10^{12}$ &  $0.76$ & $ 4\times 10^8$~ & \cref{eff-pi} \\
COMET-X ($\mu$) & $1.5\times 10^{14}$ & $1$ & \cref{mu-bkg} & \cref{acc-COMET-mu} \\
COMET-X ($\pi$) & $9\times 10^{11}$ & $1$ & $4\times 10^9$ & \cref{acc-COMET-pi} \\ 
\end{tabular} 
\label{tab:stops}
\end{ruledtabular}
\end{table}
%

%%%%%%%%%%%%%%%%%%%%%%%%%%%%%%%%%%%%%%%%%%%%%%%
\section{Projected sensitivities \label{sec:proj}}
%%%%%%%%%%%%%%%%%%%%%%%%%%%%%%%%%%%%%%%%%%%%%%%

A search for a monoenergetic positron allows for a data driven background estimate in the signal window. For a parent particle, $P$, of mass $m_P$, the energy of the positron in a decay $P\rightarrow e^+ X$ is given by 
\begin{equation}
  P_{e^+}=\frac{m_P}{2} \sqrt{\qty(1-\frac{m_{X}^2}{m_P^2}+\frac{m_{e}^2}{m_P^2})^2-\frac{4m_e^2}{m_P^2}}~.
\end{equation}
In a statistically limited search the 90\%-CL sensitivity to the branching ratio is given by 
\begin{equation}
    \label{BR-90}
     {\rm BR}_{90}(m_X)= \qty[1.28 \times \sqrt{\mu_{\rm bkg}}] \times \frac{1}{N_{P-\rm stop}}\frac{1}{\epsilon_P(P_{e^+}) }~,
\end{equation}
where $\mu_{\rm bkg}$ is the estimated background in the signal window, and $N_{P-\rm stop}$ is the number of stopped parent particles, i.e., pions or muons. 

The number of background events for the muon decay at rest search is found by taking the tree-level Michel spectrum,  $\dd \Gamma/ \dd x =2x^2(1-2x/3)$,  where $x=2E_e/m_\mu$, and multiply by the bin width, which was taken to be given by $\Delta E_e=1~{\rm MeV}$. This gives the background estimate for the $\mu^+ \rightarrow e^+ X$ search,
\begin{equation}
    \label{mu-bkg}
    \mu_{\rm bkg}(P_{\rm e^+})= N_{P-{\rm stop}}\times \frac1{\Gamma}\frac{\dd \Gamma}{\dd E_{e}} \times \epsilon_\mu(P_{e^+})\times \Delta E_e~.
\end{equation}
The acceptance $\epsilon(P_{e^+})$ is an experiment dependent quantity. 

\subsection{Mu2e-X}
For the efficiency/acceptance, $\epsilon(P_{e^+})$, we take two different functional forms motivated by Fig.~4.5 ($50\%$ nominal $B$-field) and Fig.~6.1 ($76\%$ nominal $B$-field) in Ref.~\cite{shihua_thesis}, respectively, 
\begin{align}
    \epsilon_{\mu}(P_{e^+})&=  0.25\qty(\tfrac{P_{e^{+}}-38~{\rm MeV}}{(55-38)~{\rm MeV}}) \Theta(P_{e^+}-P_{\rm thr}^\mu)~, \label{eff-mu}\\
    \epsilon_{\pi}(P_{e^+})&= 0.28\qty(\tfrac{P_{e^{+}}-55~{\rm MeV}}{(70-55)~{\rm MeV}})^{1.7} \Theta(P_{e^+}-P_{\rm thr}^\pi)~,\label{eff-pi}
\end{align}
where $P_{\rm thr}^\pi=55~{\rm MeV}$ and $P_{\rm thr}^\mu=38~{\rm MeV}$.

For the $\pi^+ \rightarrow e^+ X$ search we take $\mu_{\rm bkg} = 4\times 10^8$ \cite{shihua_thesis} in a bin of width $\Delta E_e=1~{\rm MeV}$. This background is dominantly composed of muons decaying in flight, and we take the spectrum to be flat from 55 MeV to 70 MeV.  Resulting projections for the 90\%-CL branching ratio limits are show in  \cref{fig:sensitivity}, given the inputs in \cref{inputs}. 

\subsection{COMET-X}
COMET will also have a large sample of pion and muon decays. For Phase-I the collaboration expects $1.5\times 10^{16}$ stopped muons, whereas for Phase-II they expect $1.1 \times 10^{18}$ stopped muons \cite{COMET:2018auw}. The preferred calibration tool at COMET is $\pi^+\rightarrow e^+ \nu_e$ and there is no plan to lower the $B$-field for callibration (although polarity in the transport solenoid will be modified to deliver $\mu^+$ and $\pi^+$) \cite{Kuno_chat}. We will assume that $1\%$ of the COMET beam time will be dedicated to calibration, and therefore assume $1.5 \times 10^{14}$ stopped $\mu^+$ in Phase-I and $1.1\times 10^{16}$ stopped $\mu^+$ in Phase-II. In our projections we used Phase-I, however these can be easily re-scaled to account for the increased statistics in Phase-II, or for a different fraction of runtime spent on callibration. 

We take a model for the efficiency at COMET motivated by the ``no blocker'' curve of  Fig.~7 in Ref.~\cite{Xing:2022rob}, which is well described by the following functional form, 
\begin{align}
    \begin{split}
    \label{acc-COMET-mu}
    \epsilon_{\mu}(P_{e^+}) = &0.29\qty(0.9 + \tanh\qty[\tfrac{P_{e^+} - 57~{\rm MeV}}{13~{\rm MeV}}])\\
   &\times \qty(1 + \tfrac{P_{e^+}}{725~{\rm MeV}}) \Theta(P_{e^+}-38~{\rm MeV})~,
   \end{split}\\
   \begin{split}
       \epsilon_{\pi}(P_{e^+}) = &0.29\qty(0.9 + \tanh\qty[\tfrac{P_{e^+} - 57~{\rm MeV}}{13~{\rm MeV}}])\\
   &\times \qty(1 + \tfrac{P_{e^+}}{725~{\rm MeV}}) \Theta(P_{e^+}-55~{\rm MeV})~,
   \label{acc-COMET-pi}
   \end{split}
\end{align}
The hard cut at $55~{\rm MeV}$ for $\epsilon_\pi$ is put in by hand because we expect the $\mu^+$ DIF background to rise sharply below this energy. 
For the muon decay in flight background at COMET we do not have access to the same detailed simulations performed in \cite{shihua_thesis}. In lieu of a better quantitative procedure, we simply take the estimates for the number of $\mu^+$ DIF per stopped $\pi^+$ computed in \cite{shihua_thesis} and multiply by $10$. 
The resulting projections for the 90\%-CL branching ratio limits at COMET-X Phase I are similar to Mu2e-X, see \cref{fig:sensitivity}.

\section{Conclusions and outlook \label{concl}} 
If either Mu2e or COMET uses $\mu^+$ and/or $\pi^+$ decays at rest while validating their detector response they will have access to enormous samples of both species, potentially larger than all existing datasets by orders of magnitude. As we argued in this manuscript, there is strong potential for a rich complementary physics program using this data alone, even as a purely parasitic experiment, i.e., without independent optimizations beyond the needs of the Mu2e detector response validation. 

We advocate the use of both COMET and Mu2e's validation data to search for BSM physics, and argue that their potential impact on BSM searches is sufficiently compelling to warrant dedicated analyses; see Ref.~\cite{Mu2e:2023aaa} for efforts within Mu2e to realize this goal. In particular, we have identified two decay channels that are sensitive to well-motivated BSM physics, and that can be studied using detector response validation data: $\mu^+\rightarrow e^+ X$  and $\pi^+\rightarrow e^+X$, where $X$ is a light new physics particle. Both decays result in  a monoenergetic positron. Timing information can be used to vary the $\mu^+$ vs $\pi^+$ purity of different samples \cite{Bernstein:2019fyh}.

When statistically limited sensitivity can be achieved in the $\mu^+\rightarrow e^+ X$ search, Mu2e-X can exceed both existing laboratory experiments and even astrophysical constraints by orders of magnitude. Mu2e-X or a comparable search using COMET could then serve as grounds for the discovery of a number of well motivated UV-completions. In the case of $\mu\rightarrow e X$, $X$ could be a QCD axion and dark matter candidate, whose lepton flavor violating couplings to muons and electrons offer its most promising detection prospects. This impressive reach suggests that a Mu2e $\mu^+$ run should not be viewed merely as a calibration/validation tool, but will result in a valuable data sample with BSM discovery potential. Leveraging the full power of Mu2e's statistics will ultimately demand a detailed understanding of systematic uncertainties for signal regions close to the Michel edge (necessary for $m_X\lesssim 20~{\rm MeV}$); the ultimate reach will depend on detailed analyses by both Mu2e and COMET. At larger values of $m_X$ the same search can be recast as a search for a massive $Z'$ with a dominantly invisible decay mode, for example if $Z'\rightarrow \chi \bar{\chi}$ dominates, where $\chi$ is the dark matter.

Our discussion is highly specialized to the case of two-body final states which leave a monoenergetic signal electron, since this provides an unambiguous experimental signature of new physics. It may be of interest to study the sensitivity of Mu2e for three body final states, whose positron energy spectra are continuous and which would appear as a distortion of the Michel spectrum. This is similar in spirit to previous searches carried out at PIENU, but may be more difficult at Mu2e. We note that the impressive branching ratio sensitivities that we estimate above for $\pi^+\rightarrow e^+ X$ and $\mu^+ \rightarrow e^+ X$ are encouraging. They suggest that for more challenging signals perhaps branching ratios in the $({\rm few})\times 10^{-6}$ regime may be accessible. At this level, rare decay modes such as $\pi^+ \rightarrow \mu^+ e^+e^-\nu$ (current limit of ${\rm BR} < 1.6 \times 10^{-6}$ \cite{Baranov:1991uj}), or $\mu^+ \rightarrow e^+ \chi \bar{\chi}$, may be attainable. The ability of Mu2e to achieve this level of sensitivity will depend crucially on the control of systematic uncertainties.

Even within the limited scope of two body final states, a search for $\mu^+\rightarrow e^+X$ and/or $\pi\rightarrow e^+X$ represents an extremely cost effective and impactful BSM physics program with exciting discovery prospects. We note that in the case of pions, the projections presented above suggest that Mu2e offers sensitivity to HNLs that will compete with the dedicated pion experiment PIONEER \cite{PIONEER:2022yag}. We hope that our study initiates further investigations into the untapped physics potential of both the Mu2e and COMET facilities, which will deliver unprecedented statistical samples of both muons and pions. For instance, in the $\pi^+ \rightarrow e^+ X$ search, an optimized momentum degrader to suppress the background from muon decays in flight would allow the Mu2e calibration run to push further into the as-yet untouched parameter space for HNL mixing with electron neutrinos. 

In conclusion, even operating as a purely parasitic search for new physics, Mu2e-X can push into untouched parameter space, and provide impactful limits on theoretically well motivated models of new physics in only a few weeks of data taking. Projected limits from Mu2e are expected in a forthcoming publication \cite{Mu2e:2023aaa}, and we encourage COMET to similarly study the capabilities of their facility to search for light weakly coupled BSM particles.

\textbf{ Acknowledgements}:
 This work began from a series of discussions with Shihua Huang, David Koltick, and Pavel Murat. We acknowledge their contributions at various stages of this work and thank them for helping us understand the challenges that must be overcome at an experiment such as Mu2e. We thank Michael Hedges for his work on the $\pi\rightarrow e\nu$ calibration at Mu2e.  We thank Diego Redigolo  for useful discussions and feedback regarding massless axion searches, Lorenzo Calibbi for suggestions for tau lepton searches for ALPs, and Matheus Hostert for suggestions on HNL induced CLFV couplings. We thank Matheus Hostert and Diego Redigolo for feedback on an early version of this manuscript. We thank Robert Bernstein, Stefano Miscetti, and the broader Mu2e collaboration for detailed feedback on the final version of this work and coordination with Ref.~\cite{Mu2e:2023aaa}. We thank Yoshi Kuno for communications regarding COMET, and Robert Schrock for clarifications on HNL mixing limits.

 This work was supported by the U.S.\ Department of Energy, Office of Science, Office of High Energy Physics, under Award Number DE-SC0019095.
 %and DE-SC0007884.
 This manuscript has been authored by Fermi Research Alliance, LLC under Contract No. DE-AC02-07CH11359 with the U.S. Department of Energy, Office of Science, Office of High Energy Physics. JZ acknowledges support in part by the DOE grants DE-SC0011784, DE-SC1019775, and the NSF grant OAC-2103889. RP acknowledges support from the U.S. Department of Energy, Office of Science, Office of High Energy Physics, under Award Number DE-SC0011632 and the Neutrino Theory Network Program Grant under Award Number DE-AC02-07CHI11359, and by the Walter Burke Institute for Theoretical Physics. Part of this research was performed at the Kavli Institute for Theoretical Physics which is supported in part by the National Science Foundation under Grant No. NSF PHY-1748958 and at the Aspen Center for Physics, which is supported by National Science Foundation grant PHY-1607611. This material is based upon work supported by the U.S. Department of Energy, Office of Science, Office of High Energy Physics, under Award Number DE-SC0011632.
%%%%%%%%%%%%%%%%%%
%%%%%%%%%%%%%%%%%%
%%%%%%%%%%%%%%%%%%
%%%%%%%%%%%%%%%%%%

\bibliography{biblio.bib}

\end{document}